\documentclass[utf8]{FrontiersinHarvard} 

\usepackage{url,hyperref,lineno,microtype,subcaption}
\usepackage[onehalfspacing]{setspace}

\def\keyFont{\fontsize{8}{11}\helveticabold }
\def\firstAuthorLast{Hamden {et~al.}} 
\def\Authors{Erika Hamden\,$^{1,*}$, Michael H. New\,$^{2}$, D.E. (Betsy) Pugel\,$^{3}$, Michael Liemohn\,$^{4}$, Randii Wessen\,$^{5}$, Richard Quinn\,$^{6}$, Paul Propster\,$^{5}$, Kirsten Petree\,$^{2}$, Ellen M. Gertsen$^{2}$, Paula Evans$^{7}$, and Nicole Cabrera Salazar\,$^{8}$}
\def\Address{$^{1}$Steward Observatory, University of Arizona, 933 S Cherry Ave, Tucson AZ, 85719, USA \\
$^{2}$National Aeronautics and Space Administration Headquarters, 300 Hidden Figures Way SW, Washington D.C., 20546, USA \\
$^{3}$NASA Goddard Space Flight Center, 8800 Greenbelt Rd, Greenbelt, MD 20771, USA\\
$^{4}$Department of Climate and Space Sciences and Engineering, University of Michigan, Ann Arbor, MI, 48109, USA \\
$^{5}$Jet Propulsion Laboratory, California Institute of Technology, Pasadena, CA 91109, USA \\
$^{6}$NASA Ames Research Center, Moffett Field, CA, 94035, USA \\
$^{7}$LMI Consulting, LLC, 7940 Jones Branch Dr, Tysons, VA 22102, USA \\
$^{8}$Movement Consulting, LLC, Atlanta, GA, USA}
\def\corrAuthor{Erika Hamden}

\def\corrEmail{hamden@arizona.edu}

\begin{document}
\onecolumn
\firstpage{1}

\title[The PI Launchpad]{The PI Launchpad:\\ Expanding the base of potential Principal Investigators across space sciences} 

\author[\firstAuthorLast ]{\Authors} 
\address{} 
\correspondance{} 

\extraAuth{}

\maketitle

\begin{abstract}

The PI Launchpad attempts to provide an entry level explanation of the process of space mission development for new Principal Investigators (PIs). In particular, PI launchpad has a focus on building teams, making partnerships, and science concept maturity for a space mission concept, not necessarily technical or engineering practices. Here we briefly summarize the goals of the PI Launchpad workshops and present some results from the workshops held in 2019 and 2021. The workshop attempts to describe the current process of space mission development (i.e. space-based telescopes and instrument platforms, planetary missions of all types, etc.), covering a wide range of topics that a new PI may need to successfully develop a team and write a proposal. It is not designed to replace real experience but to provide an easily accessible resource for potential PIs who seek to learn more about what it takes to submit a space mission proposal, and what the first steps to take can be. The PI Launchpad was created in response to the high barrier to entry for early career or any scientist who is unfamiliar with mission design. These barriers have been outlined in several recent papers and reports, and are called out in recent space science Decadal reports. 
\tiny
 \keyFont{ \section{Keywords:} Space Missions, Space Sciences, PI training, Inclusion, Workshops} 
\end{abstract}



\section{Introduction}

The process of successful space mission development is long, iterative, and challenging. It can also be extremely rewarding, inspiring, and even fun! Due in large part to the competitive nature of the proposal process, the behind the scenes work of developing a new mission and writing the resulting proposal can be relatively opaque. It is often a challenge for new PIs to break into this space, finding themselves behind the ball from the start, uncertain of next steps, and without adequate support and resources to move forward. These obstacles are borne out by the demographics both of PIs and Science team members for selected and proposed space missions, which tend to be both very male and very white \citep{2019Centrella}. A recent report by the National Academies of Science, Engineering, and Medicine has detailed both the problem in PI demographics and made recommendations which cover, among other things, de-mystifying and simplifying the proposal process, supporting potential PIs with training, building PI training into existing missions, and supporting underrepresented groups \citep{NAP26385}. 

Here we present the PI Launchpad, a workshop which seeks to address some of the challenges a new PI will inevitably run into when developing a mission concept for the first time and give them tools and contacts to address these challenges with an eye towards mission success. The workshop is jointly funded by the Heising-Simons Foundation and NASA. The first workshop was held in November, 2019 in Tucson, Arizona, over three days. A second workshop was held virtually in June 2021 and took place over two weeks. A third workshop is in development for July 2023 in Ann Arbor, Michigan. Additional workshops will be held every two years.

\begin{figure*}[b]
    \begin{center}
        \includegraphics[width=0.75\textwidth]{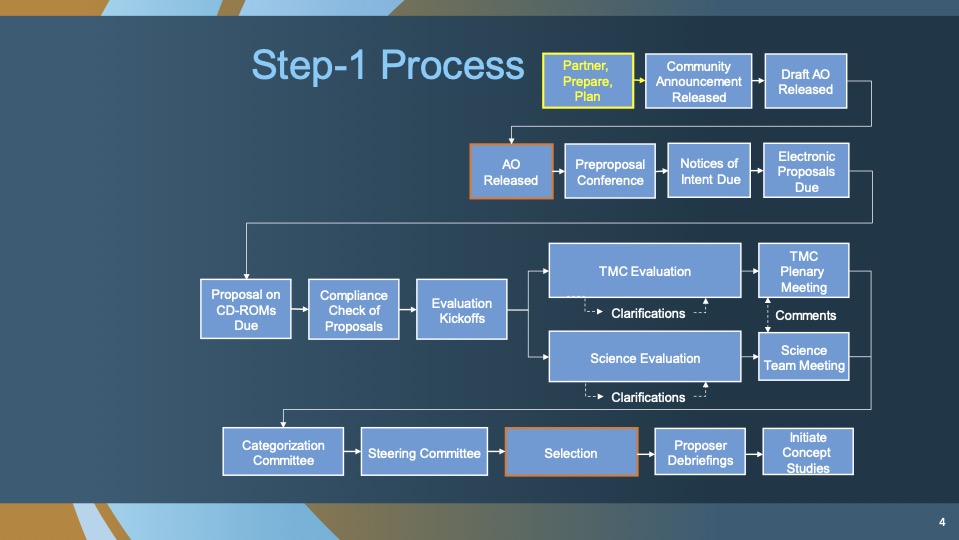}
    \end{center}
    \caption[] {\label{fig:process} A flowchart of the NASA evaluation process for PI lead missions. Slide taken from \href{https://science.nasa.gov/science-red/s3fs-public/atoms/files/PI\%20Launchpad\%20-\%20Gertsen\%20Proposal\%20Process.pdf}{2019 PI Launchpad}.}
\end{figure*}

For more detail on the proposal process itself, including NASA's review and evaluation process, please see NASA's webpage for new PIs (\href{https://science.nasa.gov/researchers/new-pi-resources}{New PI Resources}) and a colloquium by Dr. Thomas Zurbuchen which describes the evaluation process (\href{https://www.youtube.com/watch?v=xoLYRjm48-U&ab_channel=NASAVideo}{Link to Youtube}). In addition, a presentation from the 2019 PI Launchpad provides an outline for the NASA evaluation process (\href{https://science.nasa.gov/science-red/s3fs-public/atoms/files/PI\%20Launchpad\%20-\%20Gertsen\%20Proposal\%20Process.pdf}{Proposal Process}), with a graphic from this presentation shown in Figure \ref{fig:process}. For more detail on best practices for proposal development, with a focus on how to create compelling science-driven mission concepts, see \citet{2022Wessen}.

\section{The 2019 and 2021 Launchpad Workshops}
The PI Launchpad workshop addresses the challenges a new PI might face by providing information at a high level about the typical mission development process and conveying to interested scientists the recommended timelines and steps for proposing a mission.

The inaugural PI Launchpad was held in November of 2019, at the University of Arizona. It was a 2.5 day workshop and was jointly funded by NASA and through a grant from the Heising-Simons Foundation, which paid for 40 in-person attendees, supporting travel, housing, food, and transportation. The workshop was widely regarded as a success, and the participants expressed their overall enjoyment of the program captured in a report prepared by a STEM-equity consultant, Movement Consulting. Of particular note was the importance of the networking opportunities (both informal and formal) that the workshop provided. To increase outreach and accessibility, we recorded all talks and panels during the workshop, and posted closed captioned videos and materials online on a NASA-hosted website (\href{https://science.nasa.gov/researchers/pi-launchpad}{PI Launchpad}). The experience attending this workshop was described by one participant as "transformational" for them. 

A second workshop was held in July 2021, in an all-virtual format of two 90 minute sessions per day, spread out over two weeks, with a mix of panels, small group activities, lectures, and discussions. The switch to a virtual format was necessitated by the COVID-19 pandemic. There were again 40 participants. Two highlights were a panel which included all NASA Science Division Directors moderated by Prof. Erika Hamden (University of Arizona), to discuss what they were looking for in PI-led missions. This demonstrated both the buy-in from NASA decision makers for improving PI demographics, and their commitment to transparency by answering questions frankly and clearly. A second highlight was a “fireside chat” with NASA Science Associate Administrator Dr. Thomas Zurbuchen, who spoke at length with Ellen Gertsen about NASA’s overall objectives with PI-led missions. For both of these events, participants could ask questions freely of the NASA administration. In the 2021 workshop, small groups of participants were paired with a mentor virtually and there were virtual networking sessions. The virtual nature of the workshop was a hindrance in creating organic networking opportunities. As with the 2019 workshop, all content was posted online after the workshop for anyone to freely access. In addition, a report was generated by our STEM-equity consultant with suggestions for improvements and analysis of the impact of the workshop on participants.

For both workshops, the number of applications we received far exceeded the number of participants we could support. This indicates that there is still a large population of potential NASA mission PIs who want to learn the basics of building a successful proposal and team. For both workshops, we worked with a STEM Equity Consultant, Dra. Nicole Cabrera Salazar of Movement Consulting, Inc., who conducted pre- and post-workshop surveys and assessments, interviews with participants, and compiled reports which analyzed strengths and areas for improvement. Based on these reports, we know that most participants found the workshop to be incredibly valuable to them. Prior to the 2019 workshop, only 13\% of participants reported knowing what the next steps were for their mission, after the workshop, that number jumped to 90\%. For the 2021 workshop, those numbers went from 13\% who knew next steps prior to the workshop to 82\%. Anecdotally, the team has heard from numerous people, participants and non-participants, who felt that the PI Launchpad and the online content was instrumental in them developing their own mission concept. One participant e-mailed the following:
\begin{quote}
Being a part of the PI Launchpad gave me the confidence to take the reins of a project that was wildly beyond my skill set, while also giving me the tools to figure out the best way forward for my team and for my science. I'm not sure I would have agreed to be PI...without the PI Launchpad.
\end{quote}

\section{Brief overview of the PI Launchpad content}

The PI Launchpad works to cover a wide range of topics that are relevant to a new PI or mission team member. There, of course, are a nearly infinite number of topics which could be included and thus, the challenge for organizing it is to ensure that the most important topics are highlighted and given time to be explored, while also providing resources for a potential PI to continue to learn and explore the process of mission development on their own. Briefly, these topics fall into a few categories: Timelines; Mission and Science Team Roles; Developing a Science Case; Networking and Building Partnerships; Accessing Resources and Support. Both previously held workshops provided an overview of these topics, to varying levels of detail. 

\subsection{Timelines}

The time required to develop a mission concept to a level of maturity for a successful proposal varies depending roughly on the cost of the mission. An Explorer class mission (150-300 M\$) will typically take two years of development prior to submission. A concept that has already been proposed may not need as much time, since it is relying on work done previously by the mission team. A smaller mission, such as a Mission of Opportunity (MoO) or Pioneers-class mission (20-70 M\$) may only take a year of development, while larger missions such as Probes or Discovery Class missions ($>$500 M\$ to 1 B\$) may take even longer. Flagships ($>$ 1 B\$, which don't have PIs) are in process for over a decade. These are rules of thumb, and each particular experience will be slightly different. But the primary takeaway is that the earlier a PI starts their mission concept, the better positioned they will be when the Announcement of Opportunity (AO) is released and the proposal deadline is set. NASA SMD provides a projection for when they anticipate various calls for proposals coming out via the Science Office for Mission Assessments (SOMA) website (\href{https://soma.larc.nasa.gov/}{SOMA Planning Website}). At least one session per workshop is devoted to discussing timelines and how early to start. 

\begin{figure*}[t]
    \begin{center}
        \includegraphics[width=0.75\textwidth]{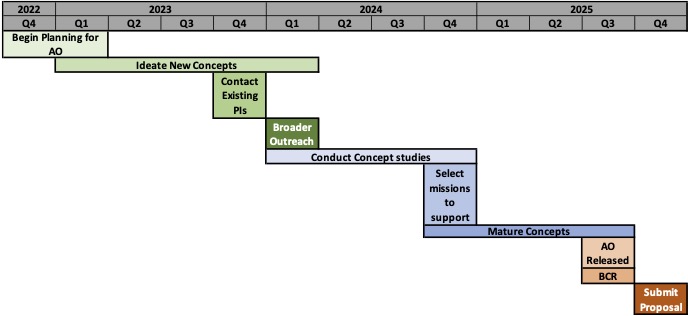}
    \end{center}
    \caption[] {\label{fig:timeline} A mock timeline for a Discovery-class mission development from a NASA center, with an expected Final AO release in Q3 of 2025. Proposal development at the center can start as early as 3 years before the expected AO. Successful missions will be in development for two years prior to the submission. This graphic was adapted from a graphic presented by JPL at the 2019 PI Launchpad (\href{https://science.nasa.gov/science-red/s3fs-public/atoms/files/PI_Launchpad_Timeline.pdf}{Timeline presentation}).}
\end{figure*}

\subsection{Mission and Science Team Roles}\label{sec:teams}

The role of the PI is just one of many critical roles in mission development. Other roles include a Deputy PI, Project Scientist, Instrument Scientist, Project Manager, Leads for various science objectives, and others. In addition, a PI will need to build a science team made up of many scientists with complimentary roles and specialities in order to ensure that the science can be achieved. Building these teams can take some time and should be approached with care. A PI needs to build a supportive team, identify key team members early, and provide team members with an understanding of the expectations for each role and the timelines involved. Team dynamics, leadership skills, and excellent communication are necessary skills for building a successful science team. Several sessions for each workshop cover topics related to this, including diversity in science teams, how to build a science team, and non-PI roles in the mission.

\subsection{Developing a Science Case}

Developing a science case is the most important aspect of building a successful mission and proposal, but it can also be one of the most challenging. A new PI may be uncertain of the maturity of their concept, uncertain if it is actually a good or a bad idea. It is a long process to turn an an initial idea into mature ``Science Objectives'' that can motivate a mission. The only way to address these concerns, mature the mission, and determine if the concept will work is to start engaging with other scientists and get additional input on possible instrument implementations. This process of development is iterative and will take time. Early on, it may feel like the science concept is too amorphous to list into objectives, or that it is difficult to achieve the level of specificity that a science objective requires. By discussing the concept with more people, and asking them to join a science team with regular meetings, a new PI can begin to hammer out what will work and what doesn't. In particular, focusing on developing a story about the science concept and building a Science Traceability Matrix (STM) can help to refine the science case. A large fraction of the PI Launchpad focuses on this, with sessions focusing on science storytelling, developing a pitch, refining science objectives, and how to use graphics to tell a story.

\subsection{Networking and Building Partnerships} 
Similar to the process highlighted in Section \ref{sec:teams}, a PI will also need to build partnerships with managing centers, industrial partners, and science team members. These partnerships take time to solidify and many potential partners will begin their process of mission development as early as 2 years before an AO will be released. This means a future PI will need to start developing a science case and then approaching possible partners between 2 and 1.5 yrs before the AO is released. The science case does not need to be finalized. In fact, it must be an iterative process that the PI conducts along with their partners. But a new PI should have an idea of what they want to explore as they begin to approach potential partners. This step can be challenging if, with new PIs frequently unsure of who to contact at a possible industrial partner. Many aerospace companies and NASA Centers have ``New Business" leads who are a good first point of contact. If they aren't the right person, they can direct a new PI to the right person. The most important step is to make an initial contact. The PI Launchpad typically has two ``Speed Networking" sessions so participants can make contacts at a range of industrial partners and NASA Centers.

\subsection{Accessing Resources and Support}
Finally, developing missions and writing proposals costs money. Many universities and institutions have funding available for a new PI, if one knows who to ask. This funding can provide partial salary support for the PI or team members, pay for engineers to create optical or mechanical designs, pay for graphics support, and additional support. In addition, institutions that have proposed missions in the past may have example proposals that can be shared with a new PI. Each institution is different, and determining what support is available is critical to secure the seed money needed for a proposal to be successful. Each workshop has at least one panel focused on what types of institutional support are available and how to access them. 

\section{Will your mission be selected?}

It is important to provide realistic expectations early on the chances of selection. Most submitted proposals are not selected. Most selected missions have been proposed multiple times. NASA's Transiting Exoplanet Survey Satellite (TESS) provides a good illustration of how long the process can be, even after a proposal is first submitted \citep{2021Ricker}. The idea for TESS came out of the team that worked on the High Energy Transient Explorer-2 (HETE-2). A group out of MIT realized that HETE-2, a UV/high-energy transient mission, could be repurposed to look for exo-planet transits, and suggested this to NASA in 2005. NASA declined, citing the upcoming launch of the Kepler mission, which was better suited to this type of work. The team then formulated the TESS concept and proposed it as a Small Explorer in 2008. It was selected for a Phase A study, but ultimately not selected for flight. The team re-proposed in 2011 as a Medium Explorer, and again was selected for a Phase A study. In this instance, it was selected for development and was launched in 2018. Thus, from first conception in 2005, TESS went through several iterations, two explorer proposal rounds, two Step 2 rounds, and 13 years of development before launch. Kepler itself was proposed 5 times before being selected in 2000 for launch in 2009. A new PI should anticipate that their experience may be similar and be prepared for multiple rounds of proposing to see an idea through.

\section{Where else to look for information}

NASA maintains a website for the PI Launchpad workshop (\href{https://science.nasa.gov/researchers/pi-launchpad}{PI Launchpad Website}), which has pdfs of presentations and recordings of many of the workshop sessions from 2019 and 2021. In addition, the PI Launchpad Workbook and additional resources are also available at the same website. The National Academies report on diversity in PI-led missions \citep{NAP26385} provides a comprehensive overview of the NASA side of the proposal evaluation process. 

\section{Discussion}

The PI Launchpad, after only two workshops, has already had a national impact. It has been directly called out in the Astro 2020 Decadal Survey \citep{2021Decadal}, and the Planetary Science and Astrobiology Decadal Survey \citep{NAP26522}, as the type of program that NASA should support and expand. The immediate impact on the 80 participants has been captured in their survey responses, but it remains to be seen what the long term impact will be. Anecdotally, many PI Launchpad participants have joined or lead science teams for proposals at all scales of mission sizes. In the long run, the impact will depend on NASA's willingness to fund and support new PIs and to require diversity in science and mission team membership. In the current status quo, compelling and groundbreaking science ideas may never see the light of day because of the challenges of being a first-time PI. We are all the poorer for it.

\section*{Conflict of Interest Statement}

Author NCS is employed by Movement Consulting, LLC and author PE was employed by NASA, but is now employed by LMI Consulting, LLC. The remaining authors declare that the research was conducted in the absence of any commercial or financial relationships that could be construed as a potential conflict of interest.

\section*{Author Contributions}

All authors provided substantial contributions to the design and execution of the PI Launchpad Workshops, either in 2019, 2021, or both. EH wrote the draft of the manuscript. All authors provided edits and comments on the draft. EH, MHN, EMG, NCS, BP, RW, PP organized the 2019 workshop. EH, MHN, BP, ML, RW, RQ, PP, KP, EMG, PE, NCS organized the 2021 workshop.

\section*{Funding}
The PI Launchpad is generously supported by a grant from the Heising-Simons Foundation. Additional support has been provided by NASA SMD.

\section*{Acknowledgments}
Some of the research was carried out at the Jet Propulsion Laboratory, California Institute of Technology, under a contract with the National Aeronautics and Space Administration. The authors acknowledge the contribution of all of the speakers and participants of the two PI launchpad workshops. In addition, we acknowledge the support and encouragement of Cyndi Atherton, of the Heising-Simons Foundation, and the support of Dr. Thomas Zurbuchen, NASA Associate Administrator for SMD.

\bibliographystyle{Frontiers-Harvard} 
\bibliography{references}

\begin{thebibliography}{6}
\providecommand{\natexlab}[1]{#1}
\expandafter\ifx\csname urlstyle\endcsname\relax
  \providecommand{\doi}[1]{doi:\discretionary{}{}{}#1}\else
  \providecommand{\doi}{doi:\discretionary{}{}{}\begingroup
  \urlstyle{rm}\Url}\fi
\providecommand{\selectlanguage}[1]{\relax}
\providecommand{\bibAnnoteFile}[1]{%
  \IfFileExists{#1}{\begin{quotation}\noindent\textsc{Key:} #1\\
  \textsc{Annotation:}\ \input{#1}\end{quotation}}{}}
\providecommand{\bibAnnote}[2]{%
  \begin{quotation}\noindent\textsc{Key:} #1\\
  \textsc{Annotation:}\ #2\end{quotation}}

\bibitem[{{Centrella} et~al.(2019){Centrella}, {New}, and
  {Thompson}}]{2019Centrella}
{Centrella}, J., {New}, M., and {Thompson}, M. (2019).
\newblock {Leadership and Participation in NASA's Astrophysics Explorer-Class
  Missions}.
\newblock In \emph{Bulletin of the American Astronomical Society}. vol.~51, 290
\bibAnnoteFile{2019Centrella}

\bibitem[{{National Academies of Sciences} and Medicine(2021)}]{2021Decadal}
{National Academies of Sciences}, E. and Medicine (2021).
\newblock \emph{{Pathways to Discovery in Astronomy and Astrophysics for the
  2020s}}.
\newblock \doi{10.17226/26141}
\bibAnnoteFile{2021Decadal}

\bibitem[{{National Academies of Sciences} and
  Medicine(2022{\natexlab{a}})}]{NAP26385}
{National Academies of Sciences}, E. and Medicine (2022{\natexlab{a}}).
\newblock \emph{Advancing Diversity Equity Inclusion, and Accessibility in the
  Leadership of Competed Space Missions} (Washington, DC: The National
  Academies Press).
\newblock \doi{10.17226/26385}
\bibAnnoteFile{NAP26385}

\bibitem[{{National Academies of Sciences} and
  Medicine(2022{\natexlab{b}})}]{NAP26522}
{National Academies of Sciences}, E. and Medicine (2022{\natexlab{b}}).
\newblock \emph{{Origins, Worlds, and Life: A Decadal Strategy for Planetary
  Science and Astrobiology 2023-2032}} (Washington, DC: The National Academies
  Press).
\newblock \doi{10.17226/26522}
\bibAnnoteFile{NAP26522}

\bibitem[{Ricker(2021)}]{2021Ricker}
Ricker, G. (2021).
\newblock Tess: A behind-the-scenes look at nasa’s latest planet hunter
  [Online; posted 25-August-2021]
\bibAnnoteFile{2021Ricker}

\bibitem[{{Wessen} et~al.(2022){Wessen}, {Propster}, {Cable}, {Case}, {Guethe},
  {Matousek} et~al.}]{2022Wessen}
{Wessen}, R.~R., {Propster}, P., {Cable}, M., {Case}, K., {Guethe}, C.,
  {Matousek}, S., et~al. (2022).
\newblock {Developing compelling and science-focused mission concepts for NASA
  competed mission proposals}.
\newblock \emph{Acta Astronautica} 191, 502--509.
\newblock \doi{10.1016/j.actaastro.2021.12.002}
\bibAnnoteFile{2022Wessen}

\end{thebibliography}


\end{document}